\documentclass[12pt]{article}
\usepackage[numbers]{natbib}
\usepackage{a4}
\usepackage{amsmath}
\usepackage{amssymb}
\usepackage{latexsym}
\usepackage{epsfig}
\def \pd{\partial}

\def \tl#1{\overset{\kern 1pt\circ}{#1}}
\def \TL#1{\overset{\kern -3pt \circ}{#1}}
\def \TLL#1{\overset{\kern -7pt \circ}{#1}}

\begin{document}
%%%%%%%%%%%%%%%%%%%%%%%%%%%%%%%%%%%%%%%%%%%%%%%%%%%%%%%%%%%%%%%%%%%%%
\title{{\bf On the Higgs mechanism and stress functions 
in the translational gauge theory of dislocations}}
\author{
Markus Lazar$^\text{}$\footnote{E-mail:
lazar@fkp.tu-darmstadt.de (M.~Lazar).} 
\\ \\
%%%${}^\text{a}$
        Emmy Noether Research Group,\\
        Department of Physics,\\
        Darmstadt University of Technology,\\
        Hochschulstr. 6,\\
        D-64289 Darmstadt, Germany\\
%%%${}^\text{b}$
%%%Institute for
%%% Theoretical Physics, University of Cologne\\ 50923 K\"oln, Germany\\
%%%${}^\text{c}$
%%%Dept.\ of Physics \& Astronomy, Univ. of Missouri-Columbia\\ Columbia, MO
%%%65211, USA
}

\date{\today}
\maketitle
%%%%%%%%%%%%%%%%%%%%%%%%%%%%%%%%%%%%%%%%%%%%%%%%%%%%%%%%%%%%%%%%%%%%%%%%%%%%%

\begin{abstract}
In this letter we discuss the Higgs mechanism in the linear and
static translational gauge theory of dislocations. We investigate the
role of the Nambu-Goldstone field and the Proca field in the
dislocation gauge theory. In addition, we give the constitutive
relations for (homogeneous) anisotropic, hemitropic and isotropic materials 
and also stress function tensors for the gauge theory of dislocations.
\end{abstract}

\section{Introduction}
The translational gauge theory of dislocations was introduced 
by~\citet{Edelen83,Edelen88} and later improved by~\citet{Lazar00,Lazar02,LA08a,LA08b}.
The gauge theory of dislocations is a continuum theory of dislocations
where the core emerges naturally making redundant the artifical cut-off radius.
Thus, in the gauge theory a dislocation is not a $\delta$-string
concentrated on the dislocation line 
with an artificial divergency like in classical dislocation theories.
Dislocations are smeared-out 
line-like objects called topological strings in gauge theories. 
In gauge theory, dislocations are not postulated a priori in an ad hoc fashion like
in classical dislocation theory rather they arise naturally as a consequence 
of broken global translation symmetry and the concept of minimal replacement.
Using local gauge invariance, the gauge theory of defects allows dislocations to 
emerge naturally and provides physical meaningful solutions (e.g., removal 
of singularities of the stress and elastic
distortion~\citep{Edelen96,Sharma05, LA08b}). 

In gauge theories the gauge bosons can be massive or massless.
Higgs fields play an important role in order to obtain
nontrivial topological structures such as Nielsen-Olesen vortices or monopoles.
On the other hand, Higgs fields provide mass for the gauge bosons.
Both cases have analogs in defect theories. 
In Ref.~\citep{Lazar02} the analogy between a screw dislocation in the gauge
theory of dislocations and the Nielsen-Olesen vortex is shown 
(see also~\citep{Gairola93}).
\citet{LA08b} derived the gauge theoretical solutions for screw and edge dislocations depending on several internal length scales. These solutions are vortex-type solutions. 
Another question is if the dislocation is a massive or massless field. 
What about the Higgs mechanism in the translational gauge theory of dislocations? 
What is the physical meaning of stress functions in the dislocation gauge theory?
What about the Higgs mechanism in the so-called stress gauge formalism.
In this letter, we try to give the corresponding answers. 
First, we start with linear asymmetric elasticity. Then we consider a pure dislocation theory
without force stresses. Finally, we investigate in the framework of
translational gauge theory of dislocations the coupled system of asymmetric 
elasticity and the dislocation gauge fields.

\section{The theory of linear asymmetric elasticity}

In linear (asymmetric) elasticity the strain energy is of the form
\begin{align}
W_{\text{el}}=\frac{1}{2}\,C_{ijkl} u_{i,j} u_{k,l},
\end{align}
where $u_{i,j}$ denotes the displacement gradient and $u_i$ is the displacement vector.
Here, the elasticity tensor $C_{ijkl}$ possesses the symmetry
\begin{align}
C_{ijkl}=C_{klij}.
\end{align}
For an anisotropic material it has, in general, 45 independent components. 
We deal with asymmetric elasticity theory since it is thus possible
to insure the existence of the inverse of the tensor $C_{ijkl}$.
The Euler-Lagrange equations of linear elasticity are
\begin{align}
E^{u}_i=\frac{\pd W_{\text{el}}}{\pd u_{i}}-\pd_j \, \frac{\pd W_{\text{el}}}{\pd u_{i,j}}=0.
\end{align}
It is the force equilibrium condition.
The force stress tensor is defined by the following constitutive relation
\begin{align}
\label{ST}
\sigma_{ij}=\frac{\pd W_{\text{el}}}{\pd u_{i,j}}=C_{ijkl} u_{k,l}
\end{align}
and the force equilibrium condition reads
\begin{align}
\label{F-EQ}
\sigma_{ij,j}=0.
\end{align} 
It should be emphasized that the force stress tensor~(\ref{ST}) is an asymmetric
tensor field.
It can be seen in Eq.~(\ref{ST}) that in asymmetric elasticity theory the 
elasticity tensor $C_{ijkl}$ is a linear transformation from the space of all 
tensors of rank two into the space of all asymmetric tensors of rank two.
For that reason the elasticity tensor is invertible in linear asymmetric elasticity 
theory unlike classical elasticity\footnote{
In classical elasticity, the elasticity 
tensor, when regarded as a linear transformation on the space of all tensors of rank 
two into the space of all symmetric tensors of rank two, cannot be invertible since
its value on every skew tensor of rank two is zero (see,
e.g.,~\citep{Gurtin}).}.
Thus, the inverse expression of~(\ref{ST}) is
\begin{align}
\label{HL-inv}
u_{i,j}=C^{-1}_{ijkl}\sigma_{kl},
\end{align}
where $C^{-1}_{ijkl}$ is the inverse tensor of $C_{ijkl}$ called the compliance tensor.
They have to fulfill the relation
\begin{align}
\label{C-1}
C^{-1}_{ijkl}C_{klmn}=\delta_{im}\delta_{jn}.
\end{align}
The compatibility condition for the displacement gradient reads
\begin{align}
\label{CC-u}
\epsilon_{jkl}u_{i,kl}=0.
\end{align}

In order to satisfy automatically Eq.~(\ref{F-EQ}), one usually introduces
a stress function tensor of first order $F_{ij}$ (see, e.g., \citep{Kroener58})
\begin{align}
\label{SF-1}
\sigma_{ij}=\epsilon_{jkl}F_{il,k},
\end{align}
which has 9 components.
Substituting Eqs.~(\ref{SF-1}) and (\ref{HL-inv}) into (\ref{CC-u}),
the stress function tensor has to fulfill the 
compatibility condition
\begin{align}
 C^{-1}_{ikmn}\epsilon_{jkl} \epsilon_{npq}F_{mp,lq}=0.
\end{align}
The stress function is not unique due to a `stress gauge' transformation  
with so-called 
null stress functions $F^0_i$:
\begin{align}
\label{SF-g}
F'_{ij}=F_{ij}-F^0_{i,j}.
\end{align}
The null stress function might be also called stress gauge field. 
%%%%(see, e.g.~\citep{Kleinert}).
If we choose the gauge condition
\begin{align}
F_{ij,j}=0,\qquad{\text{with}}\qquad F^0_{i,jj}=0,
\end{align}
3 components can be eliminated. Then the stress function tensor possesses
only 6 independent components (see also the discussion on the Coloumb gauge below).

The equilibrium condition~(\ref{F-EQ}) 
reads in terms of the displacement field:
\begin{align}
C_{ijkl} u_{k,lj}=0.
\end{align}
In the isotropic case the elasticity tensor possesses 3
independent components
\begin{align}
\label{C}
C_{ijkl}=\lambda \delta_{ij}\delta_{kl}+(\mu+\gamma) \delta_{ik}\delta_{jl}+(\mu-\gamma) \delta_{il}\delta_{jk},
\end{align}
which is an isotropic tensor of rank four.
The positive definiteness of the distortion energy ($W_{\text{el}}>0$)
requires the restriction
\begin{align}
\mu> 0,\qquad\gamma > 0,\qquad 2\mu+3\lambda> 0 .
\end{align}
The Euler-Lagrange equation for an isotropic material reads
\begin{align}
(\mu+\gamma) u_{i,ll}+(\lambda+\mu-\gamma) u_{l,li}=0,
\end{align}
which is the Navier equation for asymmetric elasticity.

Thus, the displacement vector $u_i$ is a `massless' vector field with 3 degrees of freedom.
It can be seen that there is an asymmetry between the displacement vector $u_i$ (3 components) and the stress function tensor $F_{ij}$ (6 independent components).

\section{Linear dislocation theory without sources -- a pseudomoment stress theory}
First we consider a formal dislocation theory without force stresses like 
a magnetic field theory in vacuum.
In such a translational gauge theory of dislocations, which is a pseudomoment stress 
theory, the quadratic dislocation energy is given by
\begin{align}
\label{W-core}
 W_{\rm disl}=
\frac{1}{4}\, A_{ijklmn}T_{ijk} T_{lmn}.
\end{align}
The torsion tensor (or dislocation density tensor) is defined as
follows
\begin{align}
\label{Tor} 
T_{ijk}=\phi_{ik,j}-\phi_{ij,k},\qquad
T_{ijk}=-T_{ikj},
\end{align}
where $\phi_{ij}$ is the translational gauge field (gauge boson) or 
plastic gauge field. Here $\phi_{ij}$ is an asymmetric tensor field which
possesses 9 components. 
The dislocation density tensor is the translational field strength and
plays the same role as the magnetic field strength in magnetic
field theory. Furthermore, the gauge field $\phi_{ij}$ is similar like 
the magnetic potential.
$T_{ijk}$ fulfills the translational Bianchi
identity
\begin{align}
\label{BI}
\epsilon_{jkl} T_{ijk,l}=0,
\end{align}
which is the compatibility condition for $T_{ijk}$.
It means that dislocations do not have sources.
The material tensor $A_{ijklmn}$ possesses the symmetries
\begin{align}
\label{A-sym}
A_{ijklmn}=A_{lmnijk}=-A_{ikjlmn}=-A_{ijklnm}.
\end{align}
In the anisotropic case it has 45 independent components. The
Euler-Lagrange equation with respect to the translations gauge
field is given by
\begin{align}
E^{\phi}_{ij}=\frac{\pd W_{\text{disl}}}{\pd \phi_{ij}}-\pd_k \, \frac{\pd W_{\text{disl}}}{\pd \phi_{ij,k}}=0.
\end{align}
Here, the pseudomoment stress tensor is defined as the response to
dislocations
\begin{align}
\label{moment}
H_{ijk}=2\frac{\pd{W}_{\rm disl}}{\pd T_{ijk}}=-\frac{\pd{W}_{\rm disl}}{\pd \phi_{ij,k}}=A_{ijklmn}T_{lmn},\qquad H_{ijk}=-H_{ikj}.
\end{align}
This means that at all positions where the dislocation density tensor (or torsion tensor)
is non-zero, pseudomoment stresses are present.
Then, the Euler-Lagrange equations are of the form
\begin{align}
\label{M-EQ}
H_{ijk,k}=0,
\end{align}
and explicitly in the anisotropic case they read
\begin{align}
\label{FE-T-an}
A_{ijklmn}T_{lmn,k}=0.
\end{align}
It can be seen that the pseudomoment stress does not have sources.
Thus, $\phi_{ij}$ is a `massless' tensor field.
The inverse expression of~(\ref{moment}) is given by
\begin{align}
\label{HT-inv}
T_{ijk}=A^{-1}_{ijklmn}H_{lmn},
\end{align}
where $A^{-1}_{ijklmn}$ is the inverse tensor of the tensor $A_{ijklmn}$.
Therefore, they have to fulfill the relation
\begin{align}
\label{A-1}
A^{-1}_{ijklmn}A_{lmnpqr}=\delta_{ip}\delta_{jq}\delta_{kr}.
\end{align}

In order to satisfy Eq.~(\ref{M-EQ}), one may introduce a stress function
vector $\psi_i$:
\begin{align}
\label{SF-2}
H_{ijk}=-\epsilon_{jkl}\psi_{i,l}.
\end{align}
If we substitute (\ref{SF-2}) with (\ref{moment}) into the Bianchi 
identity~(\ref{BI}), we obtain the following equation for
$\psi_i$:
\begin{align}
A^{-1}_{ijkmnp}\epsilon_{jkl} \epsilon_{npq}\psi_{m,lq}=0.
\end{align}

In addition, we want to mention that any homogeneous distribution of 
dislocations with a constant dislocation density 
%%%%$three perpendicular forests of screw dislocations of equal strength $B$, 
%%%$T_{ijk}=B\epsilon_{ijk}$ 
is a simple solution of Eq.~(\ref{FE-T-an}).
They produce no force stresses and the pseudomoment stresses are
constant.
%%%%$H_{ijk}=B\, A_{[ijk][lmn]} \epsilon_{lmn}=B\, \text{const}\, \epsilon_{ijk}$ (using the relation~(\ref{A-sym})).

For isotropic tensors of rank six we obtain in general~\citep{Jaun}
\begin{align}
\label{A-15}
A_{ijklmn}&=A_1\, \delta_{ij}\delta_{kl}\delta_{mn}+A_2\, \delta_{ij}\delta_{km}\delta_{ln}+A_3\, \delta_{ij}\delta_{kn}\delta_{lm}+
A_4\, \delta_{ik}\delta_{jl}\delta_{mn}+A_5\, \delta_{ik}\delta_{jm}\delta_{ln}
\nonumber\\
&+A_6\, \delta_{ik}\delta_{jn}\delta_{lm}+
A_7\, \delta_{il}\delta_{jk}\delta_{mn}+A_8\, \delta_{il}\delta_{jn}\delta_{km}+A_9\, \delta_{il}\delta_{jm}\delta_{kn}+
A_{10}\, \delta_{im}\delta_{jn}\delta_{kl}\nonumber\\
&+A_{11}\, \delta_{im}\delta_{jk}\delta_{ln}+A_{12}\, \delta_{im}\delta_{jl}\delta_{kn}+
A_{13}\, \delta_{in}\delta_{jm}\delta_{kl}+A_{14}\,
\delta_{in}\delta_{jl}\delta_{km}+A_{15}\, \delta_{in}\delta_{jk}\delta_{lm} ,
\end{align}
which possesses 15 independent coefficients.
If we use the first symmetry relation in Eq.~(\ref{A-sym}), we 
obtain: $A_1=A_{15}$, $A_2=A_6$, $A_4=A_{11}$, $A_{10}=A_{14}$. 
The antisymmetry in $jk$ requires: $A_2=-A_5$, $A_3=-A_6$, $A_8=-A_9$
$A_{10}=-A_{12}$, $A_{13}=-A_{14}$.
The antisymmetry in $mn$ demands: $A_2=-A_3$, $A_5=-A_6$, $A_8=-A_9$
$A_{10}=-A_{13}$, $A_{12}=-A_{14}$. 
In addition, the antisymmetries in $jk$ and $mn$ require: $A_1=0$, $A_4=0$,
$A_7=0$, $A_{11}=0$ and $A_{15}=0$.
Thus, Eq.~(\ref{A-15}) reduces to
\begin{align}
\label{A-3}
A_{ijklmn}&=A_2\big(\delta_{ij}[\delta_{km}\delta_{ln}-\delta_{kn}\delta_{lm}]
+\delta_{ik}[\delta_{jn}\delta_{lm}-\delta_{jm}\delta_{ln}]\big)
+A_8\,\delta_{il}[\delta_{jn}\delta_{km}-\delta_{jm}\delta_{kn}]\nonumber\\
&+A_{10}\big(\delta_{im}[\delta_{jn}\delta_{kl}-\delta_{jl}\delta_{kn}]
+\delta_{in}[\delta_{jl}\delta_{km}-\delta_{jm}\delta_{kl}]\big)
\end{align}
with only 3 independent coefficients.
If we introduce the abbreviations
\begin{align}
c_1=-2A_8,\qquad c_2=2A_{10},\qquad c_3=-2A_2,
\end{align}
then the pseudomoment stress~(\ref{moment}) reads (see also~\citet{LA08a,LA08b})
\begin{align}
\label{moment-iso}
H_{ijk}= c_1 T_{ijk} + c_2 (T_{jki} + T_{kij}) + c_3 (\delta_{ij}T_{llk} + \delta_{ik}T_{ljl}).
\end{align}
The positive definiteness of the dislocation energy ($W_{\text{disl}}>0$)
requires the restriction~\citep{LA08b}
\begin{align}
c_1-c_2> 0,\qquad c_1+2c_2 > 0,\qquad c_1-c_2+2c_3> 0 .
\end{align}
For the isotropic case the Euler-Lagrange equation reads
\begin{align}
\label{ME2}
&c_1(\phi_{ik,jk}- \phi_{ij,kk}) +
c_2(\phi_{ji,kk}- \phi_{jk,ki}+ \phi_{kj,ki}- \phi_{ki,kj})
\nonumber\\&\hspace{5cm}
+c_3\big[\delta_{ij}(\phi_{lk,kl}- \phi_{ll,kk})+
\phi_{ll,ji}- \phi_{lj,li}\big] =0.
%%&+ \lambda \delta_{ij}\beta_{ll} +(\mu+ \gamma)\beta_{ij}
%% + (\mu- \gamma)\beta_{ji}= \sigma^0_{ij}.
\end{align}
The gauge transformation of the three-dimensional translation
group acts in the following way on the gauge field
\begin{align}
\label{GT1}
\phi'_{ij}=\phi_{ij}-f_{i,j},
\end{align}
where $f_i$ is a vector-valued gauge function. 
The strength tensor $T_{ijk}$, the associated energy density~(\ref{W-core}) and the 
field equations are invariant under this gauge transformation.
If $\phi_{ij}$ is a solution of the field equation~(\ref{ME2}), then $\phi'_{ij}$ is 
also a solution. Thus, the general solution contains an arbitrary vector function.
This property of the non-uniqueness of a solution is inconvenient. 
Thus, a supplementary condition can be imposed on the gauge field $\phi_{ij}$, 
in order to decrease this arbitrariness and, so to speak, to fix the gauge.
For three-dimensional problems a gauge condition called Coulomb gauge is often used.
The Coulomb gauge reads
\begin{align}
\phi_{ij,j}=0.
\end{align}
This condition is not invariance under the gauge transformation~(\ref{GT1}). 
If $\phi_{ij}$ satisfies this condition, then $\phi'_{ij}$ also satisfies it,
if and only if the gauge function $f_i$ fulfills
\begin{align}
f_{i,jj}=0.
\end{align}
In this way, one can eliminate 3 (unphysical) degrees of freedom from $\phi_{ij}$.
Therefore, in the Coulomb gauge the field equation~(\ref{ME2}) for $\phi_{ij}$ simplifies to
\begin{align}
\label{ME3}
&c_1 \phi_{ij,kk} -c_2(\phi_{ji,kk}+ \phi_{kj,ki}-
\phi_{ki,kj})
%%%\nonumber\\&\hspace{5cm}
+c_3(\delta_{ij} \phi_{ll,kk}-\phi_{ll,ji}+\phi_{lj,li}) =0.
%%&+ \lambda \delta_{ij}\beta_{ll} +(\mu+ \gamma)\beta_{ij}
%% + (\mu- \gamma)\beta_{ji}= \sigma^0_{ij}.
\end{align}
If we use the Coulomb gauge, we conclude that $\phi_{ij}$ is a `massless' 
tensor field with $6=9-3$ degrees of freedom.
Thus, again there is an asymmetry between the gauge potential $\phi_{ij}$ (6 independent components) and the stress function vector $\psi_i$ (3 components).

\section{The Higgs mechanism in the translational gauge theory }
\subsection{Anisotropic case}

For an anisotropic material, if the dislocation theory is
minimally coupled with the asymmetric elasticity theory, the strain energy reads
\begin{align}
\label{W-t}
 W_{}=
\frac{1}{4}\, A_{ijklmn}T_{ijk} T_{lmn}+
\frac{1}{2}\, C_{ijkl}\nabla_j u_{i} \nabla_l  u_{k}+
B_{ijklm}\nabla_j u_i T_{klm},
\end{align}
where the translational gauge-invariant derivative is defined by
\begin{align}
\label{Gdiff} \nabla_j u_i:=u_{i,j}+\phi_{ij}.
\end{align}
The (local) gauge transformation of the three-dimensional translation group
for the displacement vector and the gauge potential are
\begin{align}
u'_i=u_i+ f_i,\qquad
\phi'_{ij}=\phi_{ij}-f_{i,j}.
\end{align}
The material tensor $B_{ijklm}$ possesses just the symmetry
\begin{align}
B_{ijklm}=-B_{ijkml}
\end{align}
and it has 81 independent components. Thus, $A_{ijklmn}$,
$B_{ijklm}$ and $C_{ijkl}$ have 171 independent components.

Instead of $u_i$ and $\phi_{ij}$ we introduce with (\ref{Gdiff}) the gauge invariant quantity
\begin{align}
\label{dist-el}
\beta_{ij}&:=\nabla_j u_i.
\end{align}
Then the dislocation density tensor can be equivalently written
\begin{align}
\label{DD-B}
T_{ijk}&=\beta_{ik,j}-\beta_{ij,k}.
\end{align}
If we use these quantities, (\ref{W-t}) takes the form
\begin{align}
\label{W-t2}
 W_{}=
\frac{1}{4}\, A_{ijklmn}T_{ijk} T_{lmn}+
\frac{1}{2}\, C_{ijkl}\beta_{ij} \beta_{kl}+
B_{ijklm}\beta_{ij} T_{klm},
\end{align}
which contains only gauge-invariant variables. 
Now some remarks
are in order. Eq.~(\ref{W-t2}) is the `Lagrangian' (strain energy) of
the `massive' field $\beta_{ij}$, which is a Proca tensor field.
The field $u_i$ does not appear in (\ref{W-t2}) et all, it needs
not to satisfy any field equation. Therefore, the massless vector
field $u_i$ is the Nambu-Goldstone vector field and $\phi_{ij}$ is
the gauge field in the translational gauge theory of
dislocations. Due to the gauge symmetry,  
the tensor field $\beta_{ij}$ has absorbed the
Nambu-Goldstone field $u_i$ and has acquired a `mass'.  
This is the essence of the Higgs mechanism.
In the theory with  the Higgs mechanism,
the tensor field $\beta_{ij}$ has 9 physical degrees of freedom, it has taken away 3
from $u_i$ and 6 from $\phi_{ij}$. No additional gauge fixing is
allowed. 
The massive tensor field appeared in a theory with gauge-invariant Lagrangian.
Thus, we conclude that the displacement field $u_i$ and the plastic gauge
field $\phi_{ij}$ are not physical state quantities, only the gauge-invariant
tensor field $\beta_{ij}$ is relevant. In dislocation theory, 
$\beta_{ij}$ is the incompatible elastic distortion tensor.
Thus, $\beta_{ij}$ and $T_{ijk}$ represent completely the geometric (or kinematic) 
degrees of freedom. 

The Euler-Lagrange equations for $\beta_{ij}$ are
\begin{align}
\label{EL-B}
E^{\beta}_{ij}=\frac{\pd W_{\text{}}}{\pd \beta_{ij}}-\pd_k \,
\frac{\pd W_{\text{}}}{\pd \beta_{ij,k}}=0.
\end{align}
With the constitutive relations for the stress tensors
\begin{align}
\label{}
\sigma_{ij}&=\frac{\pd W_{\text{}}}{\pd \beta_{ij}}=C_{ijkl} \beta_{kl}+B_{ijklm} T_{klm},\\
\label{H-const-an}
H_{ijk}&=2\frac{\pd{W}_{}}{\pd T_{ijk}}%%=-\frac{\pd{W}_{\rm disl}}{\pd \phi_{ij,k}}
=2B_{lmijk} \beta_{lm}+A_{ijklmn}T_{lmn},
\end{align}
we obtain from~(\ref{EL-B})
\begin{align}
\label{ME1}
H_{ijk,k}+\sigma_{ij}=0.
\end{align}
Thus, a translation gauge theory of dislocations is a theory with 
both force and pseudomoment stresses, where the force stress tensor is the source for
the moment stress tensor.
In the force and pseudomoment stresses model of dislocations
$\sigma_{ij}$ and $H_{ijk}$ are a complete description of the static response.

For an anisotropic material, the Euler-Lagrange equations are
\begin{align}
A_{ijklmn} T_{lmn,k}+2B_{lmijk}\beta_{lm,k}
+B_{ijklm} T_{klm}+C_{ijkl}\beta_{kl}=0
\end{align}
or in terms of $\beta_{ij}$ only
\begin{align}
A_{ijklmn}(\beta_{ln,km}-\beta_{lm,kn})+2B_{lmijk}\beta_{lm,k}
+B_{ijklm}(\beta_{km,l}-\beta_{kl,m})+C_{ijkl}\beta_{kl}=0,
\end{align}
which is gauge invariant.  
Let us mention that a gauge condition (Coulomb gauge in the static case 
or pseudo-Lorentz gauge in the dynamic case) for the gauge potentials $\phi_{ij}$ 
would destroy the gauge invariance of the Euler-Lagrange equations.
From Eq. (\ref{ME1}) we obtain the force equilibrium condition
\begin{align}
\label{FE2}
\sigma_{ij,j}=0,
\end{align}
or explicitly
\begin{align}
%%%A_{ijklmn}(\beta_{ln,km}-\beta_{lm,kn})+2B_{lmijk}\beta_{lm,k}
B_{ijklm}T_{klm,j}+C_{ijkl}\beta_{kl,j}=0
\end{align}
but $\beta_{ij,j}\neq 0$. Therefore, the force stress tensor does not have sources.

On the other hand, Eq.~(\ref{ME1}) is automatically fulfilled if we introduce 
the following stress function ansatz for the force and pseudomoment stresses:
\begin{align}
\label{SF-3}
\sigma_{ij}&=\epsilon_{jkl} F_{il,k},\\
\label{SF-4}
H_{ijk}&=-\epsilon_{jkl}(\psi_{i,l}+F_{il}).
\end{align}
We note that the translational stress gauge transformation of the stress function 
$\psi_i$ reads
\begin{align}
\psi'_i =\psi_i+F^0_i.
\end{align}
In analogy to (\ref{dist-el}), we introduce the gauge invariant stress function 
tensor
\begin{align}
\label{Psi}
\Psi_{ij}:=\tilde\nabla_j \psi_i=\psi_{i,j}+F_{ij},
\end{align} 
where $\tilde\nabla_i$ denotes the gauge-invariant derivative in the stress space.
The force stress and pseudomoment stress tensors are given then in terms 
of $\Psi_{ij}$ according to
\begin{align}
\label{SF-5}
\sigma_{ij}=\epsilon_{jkl} \Psi_{il,k},\\
\label{SF-6}
H_{ijk}=-\epsilon_{jkl}\Psi_{il}.
\end{align}
We see that the stress function tensor $\Psi_{ij}$ is equivalent to the 
pseudomoment stress tensor $H_{ijk}$. Thus, the stress function tensor $\Psi_{ij}$ is the canonical conjugate quantity to the 
dislocation density. This fact is in agreement with~\citet{Kluge,Kluge2} and 
\citet{Schaefer69}. However, constitutive relations with material tensors like 
Eq.~(\ref{H-const-an}) are missing in the description of~\citet{Kluge,Kluge2} 
and \citet{Schaefer69}.
It can be seen that 
if we use (\ref{SF-6}), then
Eq.~(\ref{SF-5}) is equivalent to (\ref{ME1}).
Also, we note that $\beta_{ij}$ and $\Psi_{ij}$ have the same number of
independent components, namely 9. $\Psi_{ij}$ has received 3 from $\psi_i$ and
6 from $F_{ij}$.
Using (\ref{H-const-an}) and (\ref{SF-6}), we can derive a field equation
for the stress function $\Psi_{ij}$.
For simplicity we just give the result for the case $B_{ijklm}=0$. 
It reads
\begin{align}
\label{SF-H}
\Psi_{it}+ A_{ijklmn}C^{-1}_{lnpq}\epsilon_{tjk} \epsilon_{qrs}\Psi_{ps,rm}=0.
\end{align}
%%%where $S_{ijkl}=C^{-1}_{ijkl}$.
In addition, the Bianchi identiy~(\ref{BI}) reads now
\begin{align}
A^{-1}_{ijkmnp}\epsilon_{jkl} \epsilon_{npq}\Psi_{mq,l}=0.
\end{align}
Since $\Psi_{ij}$ is a state quantity (pseudomoment stress) no 
gauge condition is allowed for $\Psi_{ij}$.  
Thus, the physical interpretation of Eq.~(\ref{Psi}) is the following:
$F_{ij}$ is the stress gauge field, $\psi_i$ is the stress Nambu-Goldstone 
field and $\Psi_{ij}$ is a stress Proca tensor field in a 
so-called ``stress space'' (see \citep{Kroener81} for the notation of
a stress space).
%%%which has to satisfy Eq.~(\ref{SF-H}). 
This is the Higgs mechanism in the stress gauge formalism.
The interpretation in the stress space of Eqs.~(\ref{SF-5}) and (\ref{FE2}) is the following: (\ref{SF-5}) is analogous to (\ref{DD-B}), thus it is the 
``stress gauge field strength''; (\ref{FE2}) is analogous to (\ref{BI}),
therefore, it plays the role of the ``stress Bianchi identity''.

\subsection{Isotropic non-centrosymmetric case (hemitropic materials)}
A hemitropic material is isotropic with respect to the rotation group,
however it is not invariant to coordinate inversion (non-centrosymmetric).
Sometimes such materials are called chiral materials.
For such materials the material tensors are given by (\ref{C}), (\ref{A-3}) 
and
\begin{align}
B_{ijklm}=B_{ijkn}\epsilon_{lmn},
\end{align}
where
\begin{align}
B_{ijkn}=B_1\, \delta_{ij}\delta_{kn}+B_2\, \delta_{ik}\delta_{jn}+B_3\, \delta_{in}\delta_{jk}.
\end{align}
The constitutive relations of a hemitropic theory of dislocations 
are
\begin{align}
\label{T-hem}
\sigma_{ij}&=\lambda\delta_{ij}\beta_{kk}+(\mu+\gamma)\beta_{ij}+(\mu-\gamma)\beta_{ji}+
B_{1}\delta_{ij}\epsilon_{klm} T_{klm}+B_2 \epsilon_{jlm}T_{ilm}+B_3 \epsilon_{ilm}T_{jlm},\\
\label{H-hem}
H_{ijk}&
%%%=2\frac{\pd{W}_{}}{\pd T_{ijk}}%%=-\frac{\pd{W}_{\rm disl}}{\pd \phi_{ij,k}}
=2B_{1} \epsilon_{ijk} \beta_{ll}+2B_{2} \epsilon_{jkl}\beta_{il}
+2B_{3} \epsilon_{jkl}\beta_{li}
 +c_1 T_{ijk} +c_2
 (T_{jki}+T_{kij})+c_3(\delta_{ij}T_{llk}+\delta_{ik}T_{ljl}).
\end{align}
Thus, a hemitropic material is characterized by 9 material constants.
The constants $B_1$, $B_2$ and $B_3$ are associated with noncentrosymmetry.
Finally, the Euler-Lagrange equations take the form
\begin{align}
\label{}
%%\lambda \beta_{kk,i} +
%%%(\mu + \gamma)\beta_{ij,j} + (\mu - \gamma)\beta_{ji,j}= 0\\
%%%&
&c_1 T_{ijk,k} + c_2(T_{jki,k}+T_{kij,k})
%%%\nonumber\\&\hspace{4cm}
+c_3(\delta_{ij}T_{llk,k}+T_{ljl,i})
+B_{1}\delta_{ij}\epsilon_{klm} T_{klm}
+2B_2\epsilon_{jlm}T_{ilm}+B_3
\epsilon_{ilm}T_{jlm}\nonumber\\
&\hspace{2.5cm} +2B_{1} \epsilon_{ijk} \beta_{ll,k}
%%%+2B_{2}\epsilon_{jkl}\beta_{il,k}
+2B_{3}\epsilon_{jkl}\beta_{li,k} +\lambda \delta_{ij}\beta_{ll}
+(\mu+ \gamma)\beta_{ij} + (\mu- \gamma)\beta_{ji}= 0.
\end{align}

\subsection{Chern-Simons theory of dislocations}
A special case of the hemitropic theory of dislocations is the Chern-Simons type theory of dislocations.
If  we set,
$B_1=0$ and $B_3=0$
\begin{align}
B_{ijkn}=B_2\, \delta_{ik}\delta_{jn},
\end{align}
we recover the translational Chern-Simons terms (see, e.g., Refs.~\citep{Hehl91,Mielke91}). The translational Chern-Simons theory of dislocations 
possesses 7 material constants.
The Euler-Lagrange equations take the form
\begin{align}
\label{}
%%\lambda \beta_{kk,i} +
%%%(\mu + \gamma)\beta_{ij,j} + (\mu - \gamma)\beta_{ji,j}= 0\\
%%%&
&c_1 T_{ijk,k} + c_2(T_{jki,k}+T_{kij,k})
%%%\nonumber\\&\hspace{4cm}
+c_3(\delta_{ij}T_{llk,k}+T_{ljl,i})
%%%+B_{1}\delta_{ij}\epsilon_{klm} T_{klm}
+2B_2\epsilon_{jlm}T_{ilm}
%%%%+B_3\epsilon_{ilm}T_{jlm}
\nonumber\\
&\hspace{7cm}
%%%+2B_{1} \epsilon_{ijk} \beta_{ll,k}
%%%+2B_{2}\epsilon_{jkl}\beta_{il,k}
%%+2B_{3}\epsilon_{jkl}\beta_{li,k}
+\lambda \delta_{ij}\beta_{ll}
+(\mu+ \gamma)\beta_{ij} + (\mu- \gamma)\beta_{ji}= 0.
\end{align}

\subsection{Isotropic centrosymmetric}
For an isotropic (centrosymmetric) material, we set $B_{ijklm}=0$ 
in the constitutive relations~(\ref{T-hem}) and (\ref{H-hem}).
So, the field equation for the incompatible elastic distortion 
$\beta_{ij}$ reads~\citep{LA08b}
\begin{align}
\label{FE-B}
%%\lambda \beta_{kk,i} +
%%%(\mu + \gamma)\beta_{ij,j} + (\mu - \gamma)\beta_{ji,j}= 0\\
%%%&
&c_1(\beta_{ik,jk}- \beta_{ij,kk}) +
c_2(\beta_{ji,kk}- \beta_{jk,ki}+ \beta_{kj,ki}- \beta_{ki,kj})
%%%\nonumber\\&\hspace{4cm}
+c_3\big[\delta_{ij}(\beta_{lk,kl}- \beta_{ll,kk})
+\beta_{ll,ji}- \beta_{lj,li}\big]
\nonumber\\&\hspace{7cm}
+\lambda \delta_{ij}\beta_{ll} +(\mu+ \gamma)\beta_{ij}
+ (\mu- \gamma)\beta_{ji}= 0.
\end{align}
The gauge field theoretical solutions of~(\ref{FE-B}) for 
screw and edge dislocations have been given by~\citet{LA08b}.
In Ref.~\citep{LA08b} we were able to calculate the dislocation  core for 
screw and edge dislocations. We have found that the dislocation core of an edge 
dislocation does not have a cylindrical symmetry. Also we found 
4 characteristic length scales for the translation gauge theory of dislocations.

%%%\section{Conclusion}

\section*{Acknowledgement}
%%%M.L. is supported by the DFG grant `Emmy Noether program'. 
The author has been supported by an Emmy-Noether grant of the 
Deutsche Forschungsgemeinschaft (Grant No. La1974/1-2).

\end{document}